\documentclass[11pt]{article}
\usepackage[utf8]{inputenc}
\usepackage[T1]{fontenc}
\pdfoutput = 1

\usepackage{amsmath,amssymb,amsthm}
\usepackage{graphicx,subfig}
\usepackage{changepage}
\usepackage{cite,url}
\usepackage{mathtools,xparse,MnSymbol,slashed,mathrsfs}
\usepackage{multirow}
\usepackage{stackrel}
\allowdisplaybreaks
\usepackage{makecell}

\usepackage{bbm}
\usepackage{bbold}
\usepackage{caption}
\usepackage{color}
    \definecolor{darkgreen}{rgb}{0,0.5,0}
    \definecolor{darkred}{rgb}{0.5,0,0}
    \definecolor{darkblue}{rgb}{0,0,0.6}
    \definecolor{purple}{rgb}{0.4,.2,0.7}
\usepackage{float}
\usepackage[hyperfootnotes = true, colorlinks = true, linkcolor = darkblue, citecolor = purple]{hyperref}

\numberwithin{equation}{section}

\usepackage[margin = 2.2cm]{geometry}
\usepackage[ragged]{footmisc}
\DeclareMathOperator{\Tr}{Tr}

\def\bra#1{\langle #1 |}
\def\ket#1{| #1 \rangle}
\def\inner#1#2{\langle #1 | #2 \rangle}

\usepackage[normalem]{ulem}

\begin{document}

\begin{titlepage}
\thispagestyle{empty}

\begin{flushright}
\end{flushright}

\vspace{1.2cm}
\begin{center}

\noindent{\bf \LARGE Nonperturbative Quantum Gravity in a Closed Lorentzian Universe}

\vspace{0.4cm}

{\bf \large Yasunori Nomura$^{a,b,c,d}$ and Tomonori Ugajin$^{e}$}
\vspace{0.3cm}\\

{\it $^a$ Leinweber Institute for Theoretical Physics, Department of Physics, \\
University of California, Berkeley, CA 94720, USA}\\

{\it $^b$ Theoretical Physics Group, Lawrence Berkeley National Laboratory, \\ Berkeley, CA 94720, USA}\\

{\it $^c$ RIKEN Center for Interdisciplinary Theoretical and Mathematical Sciences (iTHEMS), \\
RIKEN, Wako 351-0198, Japan}\\

{\it $^d$ Kavli Institute for the Physics and Mathematics of the Universe (WPI), \\
UTIAS, The University of Tokyo, Kashiwa, Chiba 277-8583, Japan}\\

{\it $^e$ Department of Physics, Rikkyo University, Toshima, Tokyo 171-8501, Japan}

\vspace{0.3cm}
\end{center}

\begin{abstract}
We study how meaningful physical predictions can arise in nonperturbative quantum gravity in a closed Lorentzian universe. In such settings, recent developments suggest that the quantum gravitational Hilbert space is one-dimensional and real for each $\alpha$-sector, as induced by spacetime wormholes. This appears to obstruct the conventional quantum-mechanical prescription of assigning probabilities via projection onto a basis of states. While previous approaches have introduced external observers or augmented the theory to resolve this issue, we argue that quantum gravity itself contains all the necessary ingredients to make physical predictions. We demonstrate that the emergence of classical observables and probabilistic outcomes can be understood as a consequence of partial observability:\ physical observers access only a subsystem of the universe. Tracing out the inaccessible degrees of freedom yields reduced density matrices that encode classical information, with uncertainties exponentially suppressed by the environment's entropy. We develop this perspective using both the Lorentzian path integral and operator formalisms and support it with a simple microscopic model. Our results show that quantum gravity in a closed universe naturally gives rise to meaningful, robust predictions without recourse to external constructs.
\end{abstract}

\end{titlepage}

\tableofcontents
\newpage

\section{Introduction}
\label{sec:intro}

The development of holography has revolutionized our understanding of quantum gravity. While direct applications of the AdS/CFT correspondence~\cite{Maldacena:1997re} are confined to spacetimes with boundaries, the insights gained have proven influential far beyond this limited context. In particular, holography has clarified that nonperturbative effects in gravity play a pivotal role not only at short distances near the Planck scale, but also at macroscopic distances, constraining the structure of the theory in ways not captured by conventional quantum field theory.

A striking example of this arises in closed universes, where the spacelike hypersurfaces are compact. In such settings, recent work based on the holographic entanglement entropy formula~\cite{Ryu:2006bv,Hubeny:2007xt,Faulkner:2013ana,Engelhardt:2014gca} and the gravitational path integral suggests that the nonperturbative quantum gravitational Hilbert space is one-dimensional and real~\cite{Penington:2019npb,Almheiri:2019hni,Usatyuk:2024mzs,Usatyuk:2024isz,Harlow:2023hjb} for each $\alpha$-sector~\cite{Coleman:1988cy,Giddings:1988cx,Marolf:2020xie}. This feature challenges the conventional formulation of quantum mechanics, in which predictions are made by assigning probabilities to outcomes via projection onto a complete basis of states. If the universe's state resides in a one-dimensional Hilbert space, such a procedure appears fundamentally obstructed.

To address this issue, several recent proposals introduce ``observers'' external to the system~\cite{Chandrasekaran:2022cip,Balasubramanian:2023xyd,Harlow:2025pvj,Abdalla:2025gzn,Akers:2025ahe}. These constructions effectively augment the theory with additional elements or rules not derivable from its internal dynamics, thereby specifying a basis in which projections can be meaningfully defined. While such approaches can offer practical calculational frameworks, they leave open the question of whether quantum gravity itself, as a self-contained theory, possesses the structure needed to make physical predictions.

In this paper, we argue that no such external ingredients are required. The theory of quantum gravity, when properly interpreted, contains within it all the elements needed to make meaningful predictions. Our key observation is that the emergence of classical observables and their associated probabilities can be understood as a consequence of partial observability---specifically, the fact that we do not have access to the full quantum state of the universe. Tracing out the unobserved degrees of freedom yields reduced density matrices that encode classical information and allow for robust probabilistic predictions. Importantly, this mechanism arises entirely within the framework of the original theory, without invoking projections or external observers.

We develop this viewpoint using both the gravitational path integral and operator formalisms in Lorentzian spacetimes. We show that averaging over $\alpha$-microstates---which we interpret as giving rise to effective $\alpha$-parameters at the semiclassical level---and coarse-graining over environmental degrees of freedom naturally lead to einselection and classical predictivity, with uncertainties suppressed by the entropy of the inaccessible environment.%
\footnote{
 This interpretation of $\alpha$-microstates as leading to semiclassical $\alpha$-parameters is not essential for our results, which rely only on the existence of some chaotic microscopic degrees of freedom.
}
In fact, our results suggest that it may not even be meaningful to pose physical questions about all degrees of freedom in the entire universe.

A central element of our formulation is the hierarchical structure of Hilbert spaces that emerges as we progressively impose physical constraints, moving from unconstrained auxiliary spaces to fully constrained descriptions. This hierarchy also reflects the resolution or introduction of $\alpha$-microstates, which represent internal ensemble degrees of freedom relevant to quantum gravitational processes. It provides a conceptual map of the quantum degrees of freedom and their symmetry properties at various levels. For clarity, we summarize this structure and its associated features in Table~\ref{tab:Hilbert}.
\begin{table}[t]
\begin{center}
\renewcommand{\arraystretch}{1.4}
\begin{tabular}{c|cccccccc}
  & \multicolumn{2}{c}{\raisebox{0.5\height}{unconstrained}} & \multicolumn{2}{c}{\shortstack{$d$-dim.\ constraint \\ imposed}} & \multicolumn{2}{c}{\shortstack{$D$-dim.\ constraint \\ imposed}} & \multicolumn{2}{c}{\raisebox{0.5\height}{microscopic}} \\
\hline
  Hilbert space & \multicolumn{2}{c}{${\cal H}_0 = {\cal H}_{\rm acc} \otimes {\cal H}_{\rm env}$} & \multicolumn{2}{c}{${\cal H}'_0 \simeq {\cal H}'_{\rm acc} \otimes {\cal H}'_{\rm env}$} & \multicolumn{2}{c}{$\hat{\cal H}_0 \simeq \hat{\cal H}_{\rm acc} \otimes \hat{\cal H}_{\rm env}$} & \multicolumn{2}{c}{$({\cal H}_{\rm 1d} \cong \mathbb{R}) \otimes {\cal H}_\alpha$}
\\
  Dimension & \multicolumn{2}{c}{$e^{S_{\rm univ}} \times \left( \substack{D{\rm -dim.} \\ {\rm diffeo.}} \right)$} & \multicolumn{2}{c}{$e^{S_{\rm univ}} \times \left( \substack{{\rm local\,\, time} \\ {\rm reparam.}} \right)$} & \multicolumn{2}{c}{$e^{S_{\rm univ}}$} & \multicolumn{2}{c}{$N_\alpha$}
\\
  \raisebox{-0.3\height}{\shortstack{Orthonormal \\ basis states}} & \multicolumn{2}{c}{$q({\bf x})$} & \multicolumn{2}{c}{$A$} & \multicolumn{2}{c}{$I = (n,a)$} & \multicolumn{2}{c}{$\kappa$}
\\
  \raisebox{-0.3\height}{\shortstack{Nonpert.\ quant.\ \\ grav.\ state}} & \multicolumn{2}{c}{$\rho \propto \int\!{\cal D}A\, \ket{\Psi_A} \bra{\Psi_A}$} & \multicolumn{2}{c}{$\rho \propto \int\!{\cal D}A\, \ket{\Psi_A}' \bra{\Psi_A}'$} & \multicolumn{2}{c}{$\rho \propto \mathbb{I}$} & \multicolumn{2}{c}{$\ket{\Omega_\kappa}$}
\\
  && \multicolumn{2}{c}{\hspace{1.3cm}$\underset{\text{path integral embedding}}{\xleftarrow{\hspace{2cm}}}$} & \multicolumn{2}{c}{\hspace{0.5cm}$\underset{\text{path integral embedding}}{\xleftarrow{\hspace{2cm}}}$} & \multicolumn{2}{c}{\hspace{0.5cm}$\underset{\text{tracing out ${\cal H}_\alpha$}}{\xleftarrow{\hspace{1.8cm}}}$}&
\end{tabular}
\end{center}
\caption{
 Overview of Hilbert spaces, dimensions, basis states, and representations of the nonperturbative quantum gravity state, under different stages of constraint implementation and microscopic resolution. The bottom row schematically indicates how each description is obtained from the next using path integral embedding or partial tracing. In cases where multiple spacetimes are relevant, an appropriate generalization must be implemented. Here, the $d$-dimensional constraint refers to the momentum constraint, while the $D$-dimensional one includes both the momentum and Hamiltonian constraints.
}
\label{tab:Hilbert}
\end{table}

The structure of the paper is as follows. In Section~\ref{sec:embed}, we review how quantum gravitational states satisfying the constraint equations can be embedded into the unconstrained Hilbert space using the Lorentzian path integral. We also discuss the implications of spacetime wormholes and $\alpha$-states, emphasizing the microscopic ensemble they imply. In Section~\ref{sec:pred}, we demonstrate that direct projections of the full universe state fail to yield reliable predictions due to the uniqueness of the quantum state in each $\alpha$-microstate and the existence of an ensemble of $\alpha$-microstates. We then show that tracing out the environment allows for the emergence of meaningful, stable predictions. We further illustrate this mechanism in Section~\ref{sec:operator} using a simple microscopic model. We conclude in Section~\ref{sec:concl}.

Throughout the paper, we work in a spacetime of dimension $D = d+1$, where $d$ is the number of spatial dimensions.

\section{Quantum Gravity States in Unconstrained Hilbert Space}
\label{sec:embed}

In quantum gravity, a gauge-invariant state must satisfy both the Wheeler--DeWitt~\cite{DeWitt:1967yk} and momentum constraint equations, arising from invariance under temporal and spatial diffeomorphisms, respectively. However, for many purposes---particularly when analyzing physical predictions at the semiclassical level---it is useful to consider a larger Hilbert space ${\cal H}_0$, spanned by states prior to imposing these constraints. In this section, we study how gauge-invariant states can be embedded into this enlarged Hilbert space using the gravitational path integral.

\subsection{Lorentzian gravitational path integral, group averaging, and gauged {\boldmath $CRT$}}
\label{subsec:perturb}

We begin by analyzing gauge-invariant states in perturbative quantum gravity. To obtain these states embedded in ${\cal H}_0$, we consider the Lorentzian gravitational path integral with fixed configurations for the spatial metric $h_{ij}$ and matter fields, which are collectively denoted by $\phi$, at two times $t$ and $\tilde{t}$:
\begin{equation}
  h_{ij}({\bf x},t) = h_{ij}({\bf x}),
\qquad
  \phi({\bf x},t) = \phi({\bf x})
\label{eq:bc-1}
\end{equation}
and
\begin{equation}
  h_{ij}({\bf x},\tilde{t}) = \tilde{h}_{ij}({\bf x}),
\qquad
  \phi({\bf x},\tilde{t}) = \tilde{\phi}({\bf x}),
\label{eq:bc-2}
\end{equation}
where $i,j=1,\ldots,d$. We denote the collection of spatial configurations $\tilde{h}_{ij}({\bf x})$ and $\tilde{\phi}({\bf x})$, identified under $d$-dimensional spatial diffeomorphisms of the $\tilde{t}$ hypersurface, by a label $A$, which runs continuously over the space of $d$-dimensional geometries and matter configurations (and possible boundary conditions if the hypersurface has a boundary). We denote by ${\cal H}'_0$ the Hilbert space obtained by modding out ${\cal H}_0$ by $d$-dimensional spatial diffeomorphisms; its orthonormal basis states are labeled by $A$. The resulting path integral then depends on $h_{ij}({\bf x})$ and $\phi({\bf x})$, as well as $A$.

The path integral is performed using the canonical variables in the Arnowitt--Deser--Misner (ADM) formalism~\cite{Arnowitt:1959ah}. To make the path integral well defined, the path integral measure must be divided by the gauge volume. This can be done by fixing a gauge~\cite{Teitelboim:1981ua,Halliwell:1988wc,Halliwell:1990qr}, giving
\begin{equation}
  G_A[h_{ij}({\bf x}),\phi({\bf x})] = \int\! {\cal D}h_{ij} {\cal D}\phi {\cal D}\pi^{ij} {\cal D}\pi {\cal D}N {\cal D}N^i {\cal D}\Pi\, {\cal D}\Pi_i {\cal D}c {\cal D}\bar{c} {\cal D}\rho {\cal D}\bar{\rho}\, e^{i{\cal I}},
\label{eq:G_A}
\end{equation}
where $\pi^{ij}$ and $\pi$ are the canonical momenta conjugate to $h_{ij}$ and $\phi$, respectively; $N$ and $N^i$ are the lapse function and shift vector; $\Pi$ and $\Pi_i$ are Lagrange multipliers imposing the gauge conditions; and $c$ and $\bar{c}$ are the ghosts, with $\rho$, $\bar{\rho}$ their conjugate momenta. The action is given by
\begin{equation}
  {\cal I} = {\cal I}_0 + {\cal I}_{\rm gf} + {\cal I}_{\rm gh}.
\label{eq:action}
\end{equation}
Here,
\begin{equation}
  {\cal I}_0 = \int\! d^{d+1}\!x\, [\pi^{ij} \dot{h}_{ij} + \pi \dot{\phi} - {\cal H}_G],
\end{equation}
with the time derivatives $\dot{h}_{ij}$ and $\dot{\phi}$ being implicit functions of the other variables, and ${\cal H}_G$ denoting the Hamiltonian density, which takes the form
\begin{equation}
  {\cal H}_G = \sqrt{-h} \left\{ N {\cal H} + N^i {\cal P}_i \right\},
\end{equation}
where ${\cal H}$ and ${\cal P}_i$ are the Hamiltonian and momentum constraints.%
\footnote{
 If the spatial manifold has a boundary, ${\cal H}_G$ can have additional boundary terms.
}
The gauge-fixing term takes the form
\begin{equation}
  {\cal I}_{\rm gf} = \int\! d^{d+1}\!x\, \left\{ \Pi\, [\dot{N} - \chi] + \Pi_i [N^i - \chi^i] \right\},
\end{equation}
where $\chi$ and $\chi^i$ are arbitrary functions of $h_{ij}$, $\phi$, $\pi^{ij}$, and $\pi$, and ${\cal I}_{\rm gh}$ denotes the ghost action~\cite{Fradkin:1975cq,Batalin:1983bs}. The boundary conditions are given by Eqs.~(\ref{eq:bc-1}) and~(\ref{eq:bc-2}) as well as
\begin{equation}
  \Pi({\bf x},t) = \Pi({\bf x},\tilde{t}) = \Pi_i({\bf x},t) = \Pi_i({\bf x},\tilde{t}) = c({\bf x},t) = c({\bf x},\tilde{t}) = \bar{c}({\bf x},t) = \bar{c}({\bf x},\tilde{t}) = 0,
\end{equation}
with all other variables free at the endpoints.

The quantities $G_A$ in Eq.~(\ref{eq:G_A}), when viewed as functionals of $h_{ij}({\bf x})$ and $\phi({\bf x})$, satisfy the Wheeler--DeWitt equation as well as the $d$-momentum constraints.%
\footnote{
 More precisely, one can define the path integral (e.g., via skeletonization) such that the $G_A$ satisfy the Wheeler--DeWitt equation~\cite{Halliwell:1988wc,Halliwell:1990qr}. We assume that such a definition has been adopted.
}
For this to be the case, it is important that the range of functional integration for the lapse $N$ includes negative values, which correspond to evolution backward in time. This can be seen more easily by switching to the operator formalism, in which $G_A$ can be written as the inner product between the state $\ket{h_{ij}({\bf x}),\phi({\bf x})}$ ($\in {\cal H}_0$) and the state $\ket{\Psi_A}$ obtained by evolving the state $\ket{A}$, defined in ${\cal H}_0$ modulo $d$-dimensional diffeomorphisms:%
\footnote{
 We use the symbol $\ket{A}$ to mean a representative of states that are equivalent under $d$-dimensional diffeomorphisms in ${\cal H}_0$. If we want to represent the corresponding single element of ${\cal H}'_0$, we use $\ket{A}'$. We adopt the same notation for $\ket{\Psi_A}$.
}
\begin{equation}
  \ket{\Psi_A} \sim \int\! d\tau({\bf x})\, e^{-i \int\! d^d{\bf x}\, {\cal H}({\bf x}) \tau({\bf x})} \ket{A},
\label{eq:const-imp}
\end{equation}
where $\tau({\bf x}) = N({\bf x}) (t-\tilde{t})$, and we have adopted the gauge $\chi = \chi^i = 0$. By taking the range of the functional integral to be from $-\infty$ to $\infty$, this yields the delta functional $\delta[{\cal H}({\bf x})]$, enforcing the Hamiltonian constraint. In the language of the operator formalism, the path integral of Eq.~(\ref{eq:G_A}) can thus be viewed as producing a set of (unnormalized) states $\ket{\Psi_A}$ labeled by $A$ which satisfy the constraint equations:
\begin{equation}
  \ket{\Psi_A} = \int\! {\cal D}q({\bf x})\, G_A[q({\bf x})]\, \ket{q({\bf x})},
\label{eq:Psi_A}
\end{equation}
where we have denoted field configurations $\{ h_{ij}({\bf x}), \phi({\bf x}) \}$ collectively by $q({\bf x})$; $\ket{\Psi_A}$ and $\ket{q({\bf x})}$ are elements in ${\cal H}_0$, with $\ket{q({\bf x})}$ forming an orthonormal basis of ${\cal H}_0$:\ $\inner{q({\bf x})}{q'({\bf x})} = \delta[q({\bf x}) - q'({\bf x})]$. The index $A$ of the state indicates the fact that it depends only on an equivalence class of $\tilde{q}({\bf x})$ under $d$-dimensional diffeomorphisms.

To implement the above construction, it is imperative that the path integral be performed in Lorentzian spacetime, since a well-defined analog in Euclidean spacetime does not appear to exist. In the operator formalism, the quantity $G_A$ can be viewed as the inner product~\cite{Higuchi:1991tm,Ashtekar:1995zh,Marolf:1995cn,Marolf:2000iq}
\begin{equation}
  G_A[q({\bf x})] = \bra{q({\bf x})}\,\eta\,\ket{\tilde{q}({\bf x})},
\label{eq:rel-ga}
\end{equation}
where $A$ labels field configurations $\tilde{q}({\bf x})$ modulo equivalence under $d$-dimensional diffeomorphisms. The Hermitian and positive semi-definite operator $\eta$ is called the rigging map, a map from ${\cal H}_0$ to a particular subspace $\hat{\cal H}_0^*$ of the dual Hilbert space ${\cal H}_0^*$ that satisfies the Hamiltonian and momentum constraints in the distributional sense.%
\footnote{
 While ${\cal H}'_0$ denotes the quotient of ${\cal H}_0$ under $d$-dimensional spatial diffeomorphisms, $\hat{\cal H}_0$ refers to the space of states obtained by implementing the full Hamiltonian and momentum constraints, i.e., the quotient of ${\cal H}_0$ under $D$-dimensional spacetime diffeomorphisms.
}
While we have obtained $G_A[q({\bf x})]$ via the path integral by fixing a gauge, $\eta$ can be interpreted as being given by the following formal group averaging
\begin{equation}
  \eta\, \ket{\tilde{q}({\bf x})} = \int_G\! dg\, U(g)\, \ket{\tilde{q}({\bf x})},
\end{equation}
where $U(g)$ is a unitary representation of the constraint group $G$. This perspective provides helpful intuition for later discussions. Finally, note that the distinction between bras and kets in these expressions is unimportant because of the gauged $CRT$ symmetry, which provides a natural identification between a Hilbert space and its dual, as we now see.

Since the right-hand side of Eq.~(\ref{eq:G_A}) is symmetric under the interchange of $t$ and $\tilde{t}$, 
\begin{equation}
  G_A[q({\bf x})] = \bra{q({\bf x})}\,\eta\,\ket{\tilde{q}({\bf x})} = \bra{\tilde{q}({\bf x})}\,\eta\,\ket{q({\bf x})} = (G_A[q({\bf x})])^*,
\label{eq:CRT}
\end{equation}
where we have used the Hermiticity of $\eta$. We thus find that $G_A[q({\bf x})]$---which we may also write as $G_{\tilde{A}}[\tilde{q}({\bf x})]$, with $\tilde{A}$ representing field configurations $q({\bf x})$ modulo equivalence under $d$-dimensional diffeomorphisms---is real. This is a consequence of the gauged $CRT$ symmetry~\cite{Harlow:2023hjb}, which arises automatically in a theory of quantum gravity. Note that the time evolution in Eq.~(\ref{eq:const-imp}) we have discussed here is that in the bulk, not the boundary time evolution considered in holography.

In summary, the states $\ket{\Psi_A}$ in Eq.~(\ref{eq:Psi_A}), with $G_A[q({\bf x})]$ defined by Eq.~(\ref{eq:G_A}), provide a basis for the physical Hilbert space $\hat{\cal H}_0$ embedded in ${\cal H}_0$. However, the inner products of these states cannot be defined naively in ${\cal H}_0$, since it would formally lead to an expression involving $\eta^2$, which is ill-defined~\cite{Held:2024rmg}:
\begin{align}
  \inner{\Psi_A}{\Psi_{A'}} &= \left( \int\! {\cal D}q({\bf x})\, \bra{q({\bf x})}\,\eta\,\ket{\tilde{q}({\bf x})}\, \ket{q({\bf x})} \right)^\dagger \left( \int\! {\cal D}q'({\bf x})\, \bra{q'({\bf x})}\,\eta\,\ket{\tilde{q}'({\bf x})}\, \ket{q'({\bf x})} \right)
\nonumber\\
  &= \bra{\tilde{q}({\bf x})}\,\eta^2 \ket{\tilde{q}'({\bf x})}.
\end{align}
Instead, the inner products should be defined as
\begin{equation}
  \inner{\Psi_A}{\Psi_{A'}} \equiv \bra{\tilde{q}({\bf x})}\,\eta\,\ket{\tilde{q}'({\bf x})}
  = \inner{\tilde{q}({\bf x})}{\Psi_{A'}},
\end{equation}
implying that the dual space of $\hat{\cal H}_0$ is ${\cal H}_0^*$. In the path integral, this corresponds to fixing the field configurations to be $\tilde{q}({\bf x})$ and $\tilde{q}'({\bf x})$ at the two endpoints of the time interval. Note that $A$ and $A'$ label field configurations modulo $d$-dimensional diffeomorphisms, not $D$-dimensional ones, so the inner product need not vanish even when $A \neq A'$. The set of states $\ket{\Psi_A}$ thus forms an {\it overcomplete} basis for $\hat{\cal H}_0$. This reflects the fact that the map from ${\cal H}'_0$ to $\hat{\cal H}_0$ is not injective.

\subsection{Spacetime wormholes, {\boldmath $\alpha$}-states, and an ensemble of theories}

\begin{figure}[t]
\centering
  \includegraphics[height=0.3\textwidth]{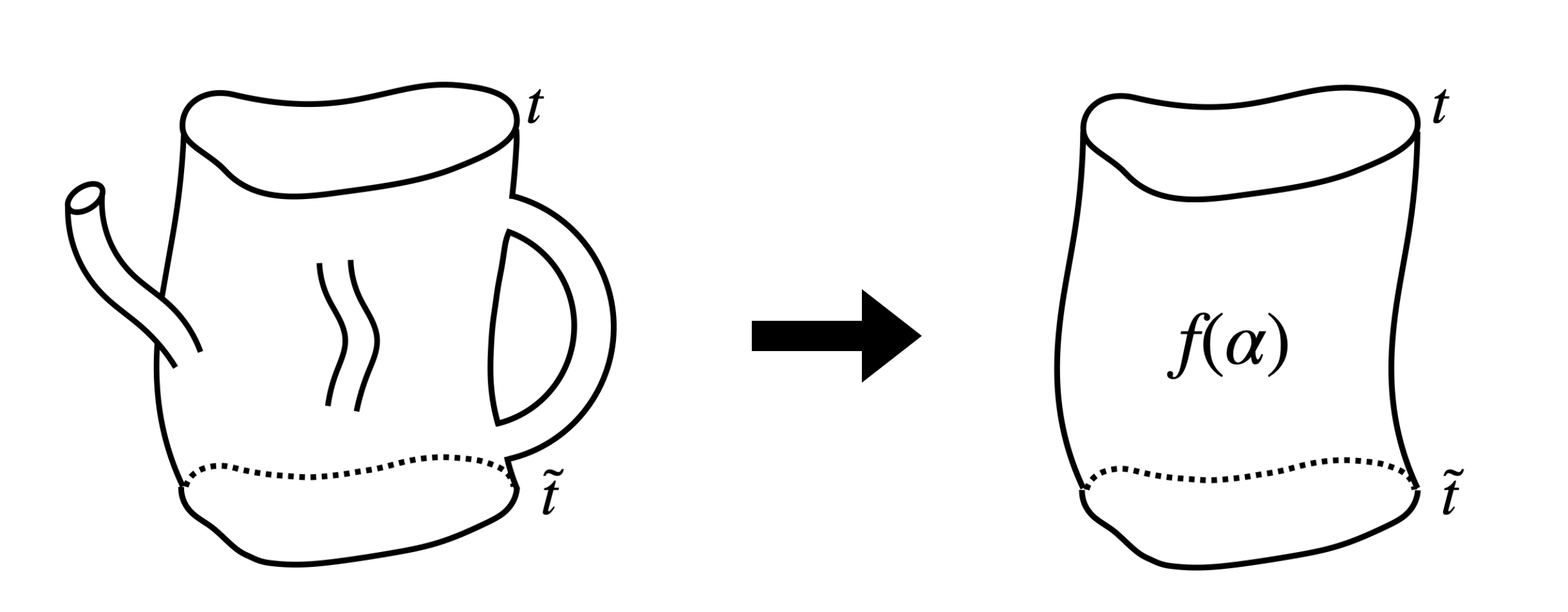}
\vspace{-0.2cm}
\caption{
 The effects of ultraviolet-scale wormholes generated via local operators can be absorbed into the probability density function $f(\alpha)$ of an ensemble of theories with varying field content, masses, and coupling constants, collectively denoted by $\alpha$.
}
\label{fig:coleman-wh}
\end{figure}

We now begin our discussion of nonperturbative effects in quantum gravity. As a first step, we explore spacetime wormholes of the type discussed in Refs.~\cite{Coleman:1988cy,Giddings:1988cx,Marolf:2020xie}. In Lorentzian spacetime, such wormholes can be incorporated by allowing certain conical-type singularities---called crotch singularities~\cite{Louko:1995jw}---to appear in the path integral. An example of such a configuration is shown in the left panel of Fig.~\ref{fig:coleman-wh}.

As discussed in Refs.~\cite{Coleman:1988cy,Giddings:1988cx}, the effects of these wormholes can be incorporated through the so-called $\alpha$-states:\ a set of states that define superselection sectors for observables within a single universe. In our context, this amounts to considering an ensemble of $D$-dimensional effective field theories with varying field content, masses, and coupling constants, which we collectively denote by $\alpha$. Since the dominant contributions from the wormholes under consideration come from small, ultraviolet-scale wormholes generated via local operators, their effects can be represented by a probability density function $f(\alpha)$ defined at the cutoff scale, as shown in the right panel of Fig.~\ref{fig:coleman-wh}. This probability distribution may also encode all the effects of physics occurring above the cutoff scale of the $D$-dimensional field theories, such as the string landscape arising from extra dimensions~\cite{Bousso:2000xa,Susskind:2003kw,Douglas:2003um}.

The fact that $\alpha$ forms a continuous set suggests that predicting the outcome of any physical measurement performed during a finite time interval involves averaging over underlying microscopic degrees of freedom not visible at the semiclassical level, but which correspond to $\alpha$ at the classical level.%
\footnote{This is because a finite-duration experiment cannot resolve a continuous set of microstates with infinite precision.
}
We assume that such averaging is implicitly incorporated in the gravitational path integral performed over semiclassical degrees of freedom, as a coarse-grained effect of integrating out microscopic degrees of freedom.%
\footnote{
 This is reminiscent of the case of a dynamical black hole, where a classical parameter such as the mass corresponds to underlying microstates that are not directly visible at the semiclassical level and are expected to behave chaotically under the microscopic dynamics~\cite{Nomura:2018kia}.
}
While we will assume below that these microscopic degrees of freedom give rise to the effective $\alpha$ parameters, this assumption is not essential for the results of the paper---the existence of some chaotic microscopic degrees of freedom is sufficient. Note that while the distribution $f(\alpha)$ cannot be determined within the effective field theory, its form is irrelevant for our purposes, as we consider an ``infinitesimal'' range of $\alpha$ corresponding to the microscopic degrees of freedom.

\subsection{Replica wormholes and the Hilbert space of a closed universe}
\label{subsec:replica}

A dramatic nonperturbative effect in quantum gravity arises when the spacelike hypersurfaces at fixed $t$ are topologically compact. In this case, the Hilbert space of states for a fixed $\alpha$-microstate is one-dimensional and real, as suggested by considerations based on the holographic entanglement entropy formula and the gravitational path integral~\cite{Penington:2019npb,Almheiri:2019hni,Usatyuk:2024mzs,Usatyuk:2024isz}.

\begin{figure}[t]
\centering
  \includegraphics[height=0.3\textwidth]{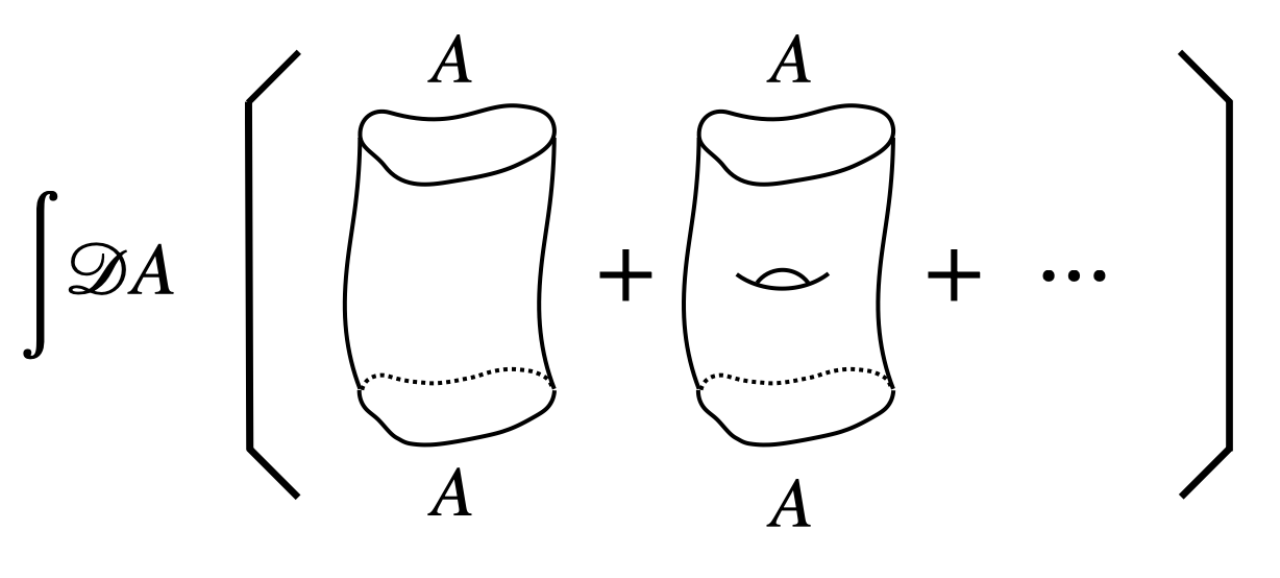}
\vspace{-0.5cm}
\caption{
 Configurations summed over in the gravitational path integral that computes $\Tr M$. Note that the lapse is integrated over both positive and negative values, corresponding to forward and backward time evolution.
}
\label{fig:Tr_M}
\end{figure}

\begin{figure}[t]
\centering
\vspace{0.5cm}
  \includegraphics[height=0.18\textwidth]{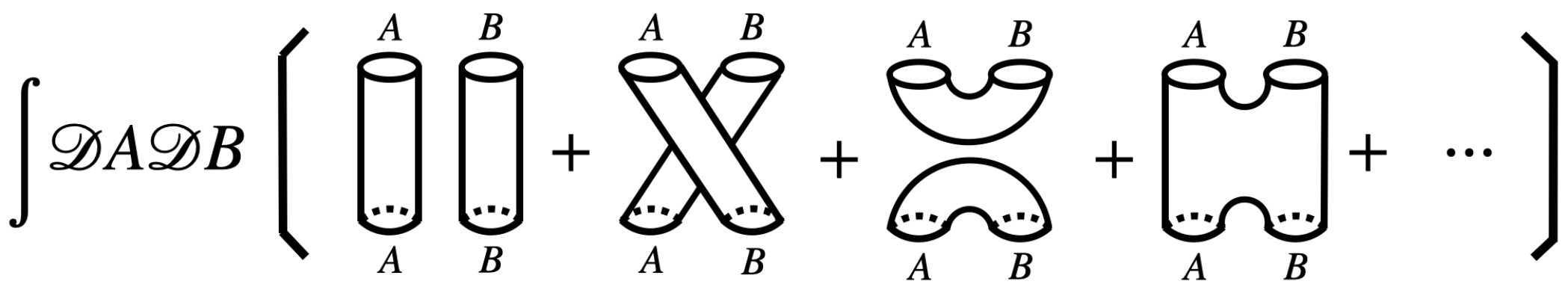}
\vspace{-0.1cm}
\caption{
 Configurations contributing to the gravitational path integral computing $(\Tr M)^2$, whose result differs from the square of $\Tr M$ due to replica wormhole contributions.
}
\label{fig:replica-wh}
\end{figure}

A key idea is that, due to the mandatory averaging procedure discussed in the previous subsection, the gravitational path integral of the product of two quantities yields a result different from the product of the gravitational path integrals of each quantity individually. For example, defining the Gram matrix
\begin{equation}
  M_{AA'} = \inner{\Psi_A}{\Psi_{A'}},
\label{eq:Gram}
\end{equation}
the quantities $\Tr M$ and $(\Tr M)^2$ can be computed by summing over configurations depicted in Fig.~\ref{fig:Tr_M} and Fig.~\ref{fig:replica-wh}, respectively.
Due to contributions from the second, third, and fourth terms (as well as analogous terms involving higher topologies) in Fig.~\ref{fig:replica-wh}---i.e., from replica wormholes~\cite{Penington:2019kki,Almheiri:2019qdq}---the result for $(\Tr M)^2$ differs from the square of $\Tr M$. This occurs because
\begin{equation}
  \overline{(\Tr M)^2} \neq \left( \overline{\Tr M} \right)^2,
\label{eq:ensemble}
\end{equation}
where overlines denote the averaging over $\alpha$-microstates discussed in the previous subsection.

In fact, the configurations summed over in Fig.~\ref{fig:replica-wh} are identical to those needed to compute $\Tr(M^2)$. This follows from the permutation invariance of the geometries summed over~\cite{Harlow:2025pvj,Abdalla:2025gzn,Balasubramanian:2025jeu}, which implies that
\begin{equation}
  \overline{\Tr(M^n)} = \overline{(\Tr M)^n}
\end{equation}
for all $n \in \mathbb{N}$. Moreover, a similar analysis shows that the variance of this relation vanishes:
\begin{equation}
  \overline{(\Tr(M^n) - (\Tr M)^n)^2} = 0.
\end{equation}
Thus, the Gram matrix satisfies
\begin{equation}
  \Tr(M^n) = (\Tr M)^n
\end{equation}
for each member of the ensemble. Since the only matrices satisfying this relation are rank-one matrices, this implies that the Hilbert space {\it for each $\alpha$-microstate} is one-dimensional. The gauged $CRT$ symmetry further implies that these Hilbert spaces are real.

Although each $\alpha$-microstate has a trivial (i.e., one-dimensional and real) Hilbert space, Eq.~(\ref{eq:ensemble}) shows that the ensemble as a whole is nontrivial, meaning it consists of multiple microstates. This feature will be important in our discussion of measurements in the next section.

\section{Emergent Predictions from Partial Observability}
\label{sec:pred}

In the remainder of this paper, we focus on the case of a closed universe, where the spacelike hypersurfaces at fixed $t$ are compact. In this case, given that the nonperturbative quantum gravitational Hilbert space is one-dimensional for each $\alpha$-microstate, it may seem possible to make predictions for any physical measurement in the form of conditional probabilities~\cite{Nomura:2011dt}, by projecting onto appropriate states representing particular outcomes of the measurement. In the present context, we might consider projecting the nonperturbative quantum gravity state onto a set of states $\ket{X_i}$ ($i = 1,2,\cdots$) in ${\cal H}_0$, corresponding to configurations in which the measurement has yielded result $X$.%
\footnote{
 These states are generally superpositions of field basis states $\ket{q({\bf x})}$.
}
The unnormalized probability of obtaining outcome $X$ would then be given by
\begin{equation}
  P(X) = \sum_i \int\!{\cal D}A\, \inner{\Psi_A}{X_i} \inner{X_i}{\Psi_A},
\label{eq:naive}
\end{equation}
which can be normalized by dividing by the sum over all possible outcomes, $\sum_X P(X)$. Here, the integral over $A$ arises because the quantum gravitational Hilbert space is one-dimensional, so the relevant quantity is the trace of the Gram matrix $M_{A A'}$ in Eq.~(\ref{eq:Gram}).

There are important virtues in the expression in Eq.~(\ref{eq:naive}). First, the fact that the measurement can occur at any time is automatically accounted for by the integration over the lapse function in preparing the state $\ket{\Psi_A}$. Second, since the constraints of general relativity are already incorporated in the construction of $\ket{\Psi_A}$, one can adopt any fixed spatial coordinates ${\bf x}$ on any fixed time slice to represent field configurations $q({\bf x})$ corresponding to $X_i$. The states $\ket{X_i}$  in Eq.~(\ref{eq:naive}) live in the unconstrained, and hence nongravitational, Hilbert space ${\cal H}_0$.

Despite these virtues, in this section we will see that the prescription in Eq.~(\ref{eq:naive}) does not work in its current form. This is partly because of the existence of $\alpha$-microstates, which introduce intrinsic uncertainties into predictions made by using Eq.~(\ref{eq:naive}), and partly because the quantum gravitational Hilbert space is one-dimensional for each $\alpha$-microstate. These issues, however, are resolved by the fact that we do not have complete knowledge of the entire universe. This allows us to make physical predictions up to an error suppressed exponentially in the entropy of the environment, i.e., the parts of the universe about which we lack information.

\subsection{Breakdown of prediction via universe state projections}
\label{subsec:breakdown}

A major conceptual problem with Eq.~(\ref{eq:naive}) arises from the fact that the nonperturbative quantum gravitational state of the universe is unique for each $\alpha$-microstate~\cite{Usatyuk:2024isz}, so this prescription necessarily involves projection operators. Normally, such projections are interpreted as reflecting the presence of an external observer, which selects the basis onto which the projection is made. In a closed universe, however, no such external observer exists. This implies that the choice of the states $\ket{X_i}$ must be specified by hand, from outside the theory. We want this basis to be determined from within the theory itself.

\begin{figure}[t]
\centering
  \includegraphics[height=0.25\textwidth]{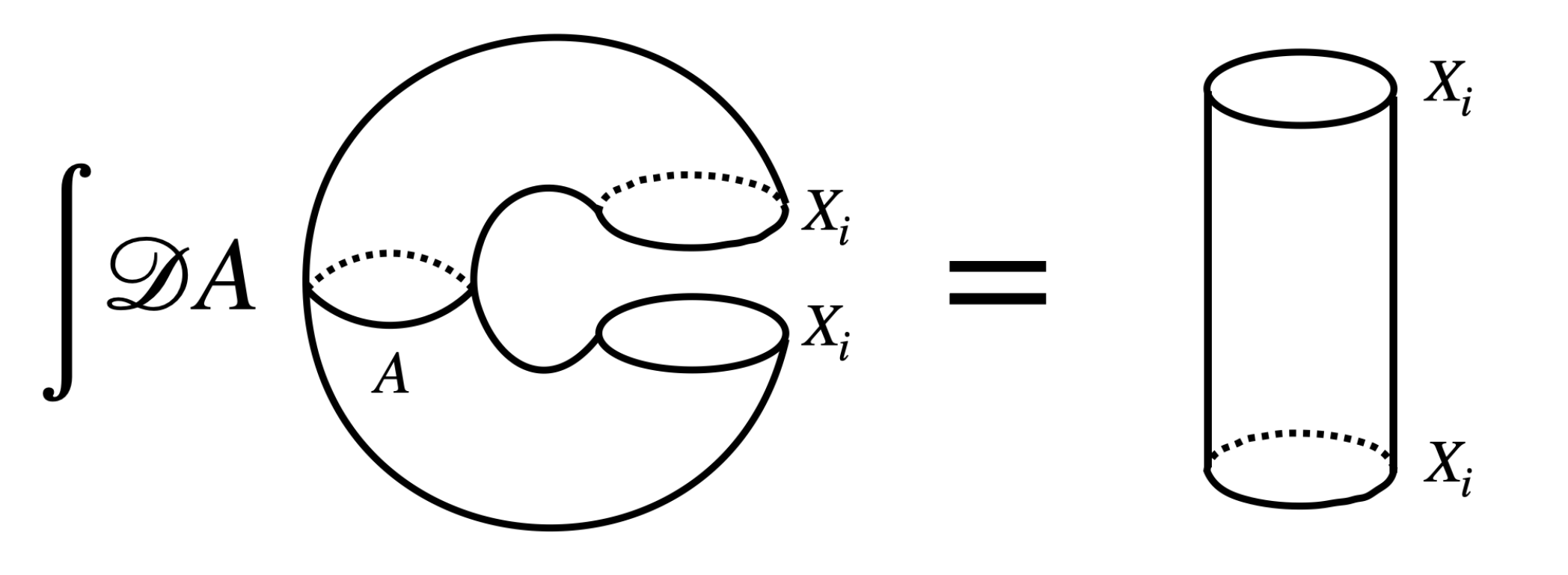}
\vspace{-0.2cm}
\caption{
 Leading-order Lorentzian path integral computing the unnormalized probability $P(X_i)$ via projection onto $\ket{X_i}$.
}
\label{fig:PXi}
\end{figure}

Moreover, even if we accept the usage of projection operators, the prescription in Eq.~(\ref{eq:naive}) still does not work. To see this, let us consider the path integral that yields
\begin{equation}
  P(X_i) = \int\!{\cal D}A\, \inner{X_i}{\Psi_A} \inner{\Psi_A}{X_i}.
\end{equation}
The leading-order configurations contributing to this are depicted in Fig.~\ref{fig:PXi}.
We can understand this in the operator formalism as
\begin{equation}
  P(X_i) = \int\!{\cal D}A\, \bra{X_i} \eta \ket{A} \inner{\Psi_A}{X_i}
  \propto \bra{X_i} \eta \ket{X_i}.
\label{eq:PXi}
\end{equation}
Here, we have used the fact that in the gravitational regime, the completeness relation takes the form
\begin{equation}
  \mathbb{I} \propto \int\!{\cal D}A\, \ket{A} \bra{\Psi_A} = \int\!{\cal D}A\, \ket{\Psi_A} \bra{A},
\end{equation}
where the proportionality factor is a state-independent constant. This reflects the fact that the dual of the constrained Hilbert space is the unconstrained Hilbert space.

\begin{figure}[t]
\centering
\vspace{0.1cm}
  \includegraphics[height=0.2\textwidth]{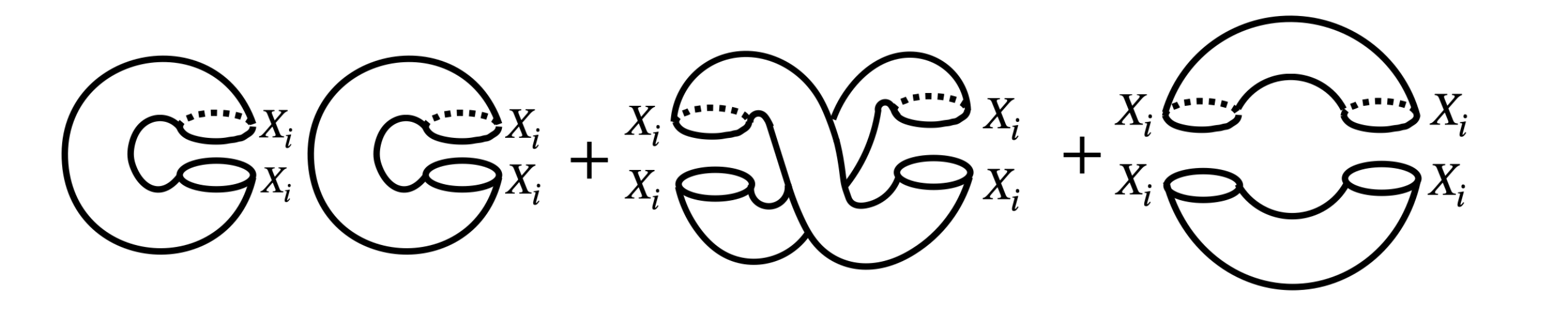}
\vspace{-0.3cm}
\caption{
 Leading-order Lorentzian path integral contributions to $P(X_i)^2$, demonstrating its non-factorization due to averaging over $\alpha$-microstates.
}
\label{fig:PXi2}
\end{figure}

The problem is that when we compute $P(X_i)^2$ using the path integral, there are three leading-order contributions, depicted in Fig.~\ref{fig:PXi2}.
As a result, the path integral does not yield the square of the value obtained from Fig.~\ref{fig:PXi}. Again, this is because the path integrals yield quantities averaged over $\alpha$-microstates, reflecting the fact that in a closed universe, we cannot determine the exact $\alpha$-microstate we inhabit.%
\footnote{
 This contrasts with setups that include an asymptotic boundary, where a hypothetical observer can, in principle, carry out an infinitely precise measurement. In such cases, it has been conjectured that including all possible physical processes allows transitions among different $\alpha$-microstates, implying that the Hilbert space for baby universes is one-dimensional~\cite{Marolf:2020xie,McNamara:2020uza}.
}
In fact, the three contributions in Fig.~\ref{fig:PXi2} are identical, so we find
\begin{equation}
  \overline{P(X_i)^2} = 3\, \overline{P(X_i)}^2,
\end{equation}
indicating that the variance of $P(X_i)$ over the ensemble of $\alpha$-microstates is large~\cite{Harlow:2025pvj}:
\begin{equation}
  \overline{(P(X_i) - \overline{P(X_i)})^2} = 2\, \overline{P(X_i)}^2.
\label{eq:PXi_Var}
\end{equation}
This implies that the actual value of $P(X_i)$ depends significantly---likely erratically---on the $\alpha$ microstate. As a result, meaningful predictions cannot be made without knowing the $\alpha$ microstate exactly.

\subsection{Stable predictions from tracing out the environment}
\label{subsec:stable}

In the discussion so far, we have implicitly treated the universe as if its full state were accessible to us, since the earlier prescription (e.g., Eq.~(\ref{eq:naive})) involved projection onto pure states. However, in any realistic situation, we can access only a portion of the entire universe. This portion contains both the system being measured and the physical observer performing the measurement. In fact, it can include all systems and observers present in the relevant branches of the wavefunction.

Let us, therefore, divide the entire universe into this accessible portion and the rest, which we call the environment. The unconstrained Hilbert space can then be factored as
\begin{equation}
  {\cal H}_0 = {\cal H}_{\rm acc} \otimes {\cal H}_{\rm env},
\label{eq:factor}
\end{equation}
where ${\cal H}_{\rm acc}$ and ${\cal H}_{\rm env}$ represent the Hilbert spaces for the accessible and the other parts of the universe.%
\footnote{
In general, the Hilbert space can be decomposed as ${\cal H}_0 \simeq \oplus_{\cal M} {\cal H}_{{\cal M},0}$ with ${\cal H}_{{\cal M},0} = {\cal H}_{{\cal M},{\rm acc}} \otimes {\cal H}_{{\cal M},{\rm env}}$, where ${\cal M}$ represents spacetimes that contribute to the path integral as saddle points. We will assume that the relevant Hilbert space is dominated by single ${\cal M}$, either because it gives the largest contribution or because we focus on a particular ${\cal M}$ selected by the physical question one asks in Eq.~(\ref{eq:pred}), as relevant in the context of environmental selection~\cite{Bousso:2000xa,Weinberg:1987dv,Vilenkin:1994ua,Hall:2007ja}.
}
While not necessary, we will, for concreteness, associate ${\cal H}_{\rm acc}$ with a subregion of the $d$-dimensional space on which the measurement conditions are imposed, and ${\cal H}_{\rm env}$ with its complement.%
\footnote{
 Strictly speaking, this identification is unrealistic in practice, as it is nearly impossible to have complete knowledge of all degrees of freedom---including microscopic ones---within a given spatial region.
}
A similar setup, involving reduced density matrices associated with spatial subregions, has been discussed in Refs.~\cite{Hartle:2016tpo,Ivo:2024ill}.

\begin{figure}[t]
\centering
\vspace{0.2cm}
  \includegraphics[height=0.3\textwidth]{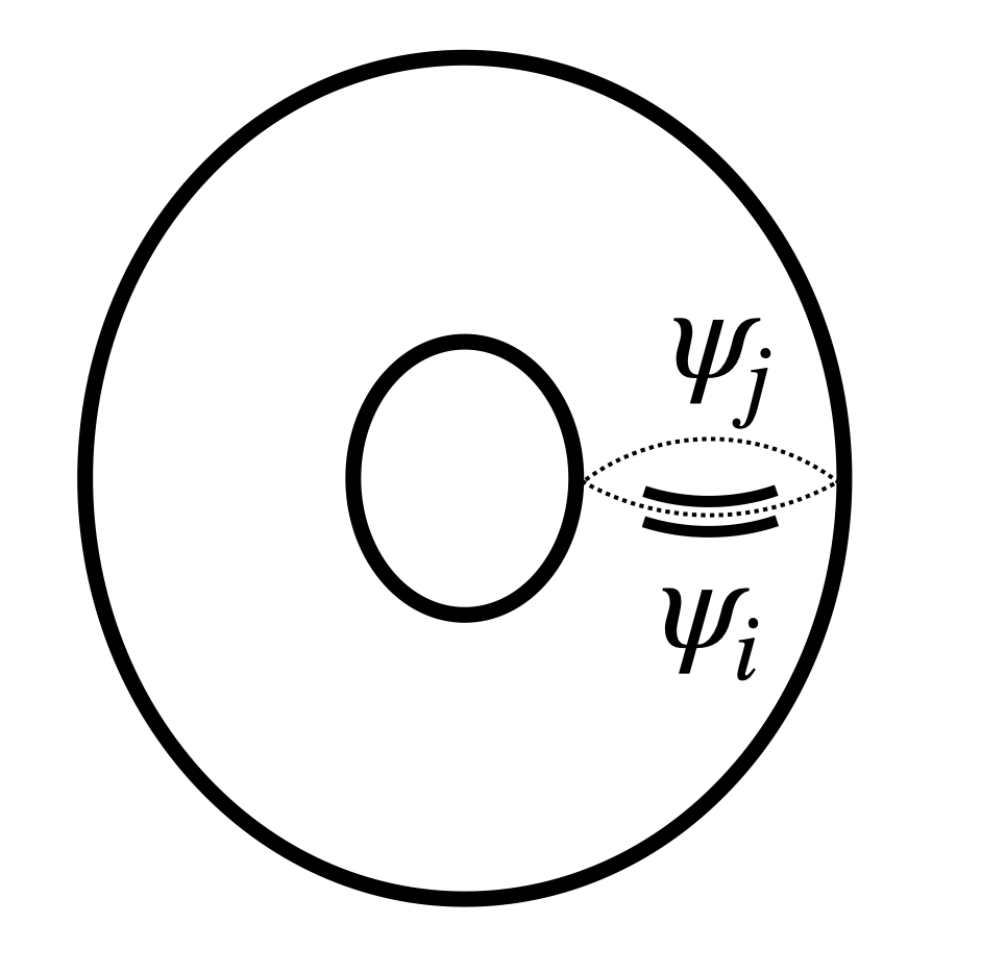}
\vspace{-0.2cm}
\caption{
 Path integral in Lorentzian spacetime computing the matrix element $\rho^{\rm acc}_{ij}$ of the reduced density operator, obtained by tracing out the environment in the unconstrained Hilbert space.
}
\label{fig:rho-acc}
\end{figure}

The (unnormalized) density operator for the entire universe is given by
\begin{equation}
  \rho = \int\! {\cal D}q({\bf x})\, {\cal D}q'({\bf x})\, \ket{q({\bf x})} \left( \int\!{\cal D}A\, \inner{q({\bf x})}{\Psi_A} \inner{\Psi_A}{q'({\bf x})} \right) \bra{q'({\bf x})}.
\end{equation}
The corresponding reduced density operator in the accessible Hilbert space ${\cal H}_{\rm acc}$ is then obtained by tracing over ${\cal H}_{\rm env}$:
\begin{align}
  \rho^{\rm acc} &= \mathop{\Tr}_{\rm env}\, \rho 
\nonumber\\
  &= \int\!{\cal D}\phi({\bf x})\! \int\!{\cal D}q({\bf x})\! \int\!{\cal D}q'({\bf x})\,\, {}_{\rm env}\inner{\phi({\bf x})}{q({\bf x})} \left( \int\!{\cal D}A\, \inner{q({\bf x})}{\Psi_A} \inner{\Psi_A}{q'({\bf x})} \right) \inner{q'({\bf x})}{\phi({\bf x})}_{\rm env},
\end{align}
where $\ket{\phi({\bf x})}_{\rm env}$ is an element of ${\cal H}_{\rm env}$, and $q({\bf x})$ and $q'({\bf x})$ include both accessible and environmental degrees of freedom. While the expression involves an integral over $A$, only the unique nonperturbative quantum gravitational state associated with each $\alpha$-microstate is physically relevant here, which we may refer to as the Lorentzian no-boundary state.%
\footnote{
 This differs from the setup in Ref.~\cite{DiazDorronsoro:2017hti}, which considers a big-bang singularity as one of the endpoints. In our case, the uniqueness of the state arises from nonperturbative effects in quantum gravity.
}
 The path integral configuration that computes the matrix element of this reduced density operator,
\begin{equation}
  \rho^{\rm acc}_{ij} = {}_{\rm acc}\bra{\psi_i({\bf x})} \rho^{\rm acc} \ket{\psi_j({\bf x})}_{\rm acc},
\end{equation}
is shown in Fig.~\ref{fig:rho-acc}.

\begin{figure}[t]
\centering
\vspace{-0.2cm}
  \includegraphics[height=0.23\textwidth]{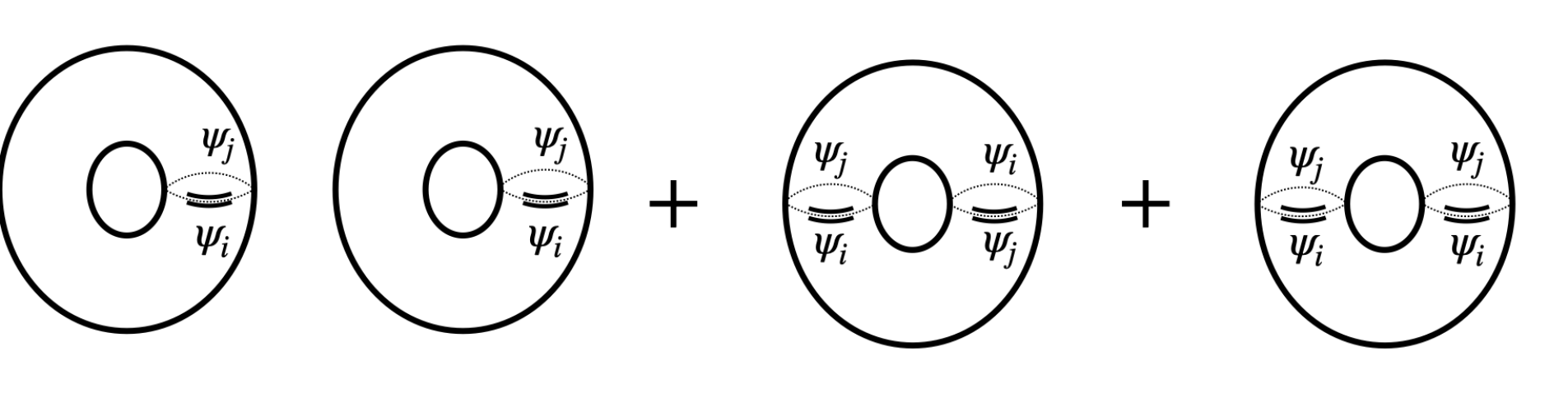}
\vspace{-0.2cm}
\caption{
 Leading contributions to $(\rho^{\rm acc}_{ij})^2$ from the Lorentzian path integral. No wormholes arise from the tracing out environmental degrees of freedom. As a result, replica wormhole terms are exponentially suppressed by the environment's entropy.
}
\label{fig:rho-acc-2}
\end{figure}

We now consider the square of the matrix element $\rho^{\rm acc}_{ij}$, as computed by the path integral. The key point is that, since the trace over ${\cal H}_{\rm env}$ is performed in the unconstrained (i.e., nongravitational) Hilbert space ${\cal H}_0$, there are no wormholes sewing together the environmental states $\ket{\phi({\bf x})}_{\rm env}$. The path integral configurations contributing to $(\rho^{\rm acc}_{ij})^2$ are shown in Fig.~\ref{fig:rho-acc-2}, where the second and third diagrams correspond to those in Fig.~\ref{fig:PXi2} (appropriately expended to all elements including off-diagonal ones).
The contributions from these diagrams are qualitatively different from those of the first term, which gives the square of the value obtained from Fig.~\ref{fig:rho-acc}. In fact, we will now see that the former is exponentially suppressed relative to the latter, with the suppression governed by the entropy of the environment.

\subsection{Emergent physical predictions and their precision in a closed universe}
\label{subsec:precision}

The second and third diagrams of Fig.~\ref{fig:rho-acc-2} represent the correction responsible for the deviation of $\overline{(\rho^{\rm acc}_{ij})^2}$ from $\overline{\rho^{\rm acc}_{ij}}^2$. Let us now estimate the size of this correction.

\begin{figure}[t]
\centering
\vspace{0.2cm}
  \includegraphics[height=0.3\textwidth]{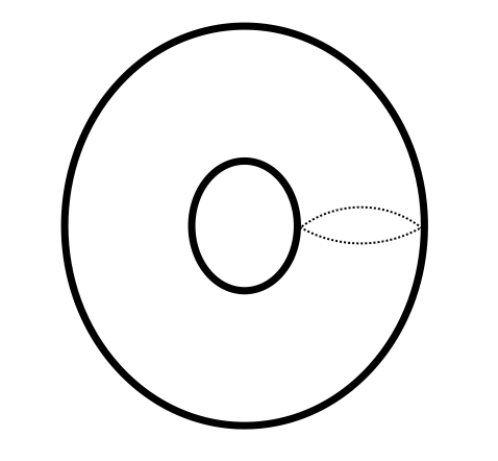}
\vspace{-0.2cm}
\caption{
 Lorentzian path integral computing $\Tr \rho^{\rm acc}$ by tracing out all the degrees of freedom in the unconstrained Hilbert space.
}
\label{fig:rho-acc-Tr}
\end{figure}

To evaluate it, we normalize $\rho^{\rm acc}$ to have unit trace. At leading order, this is achieved by dividing the contribution of the diagram in Fig.~\ref{fig:rho-acc} by that of Fig.~\ref{fig:rho-acc-Tr}, which computes $\overline{\Tr \rho^{\rm acc}}$.
Because the gluing in ${\cal H}_0$ eliminates contributions from replica wormholes, the variance of the trace vanishes at leading order, implying
\begin{equation}
  \overline{\Tr \rho^{\rm acc}} = \Tr \rho^{\rm acc} \equiv Z,
\end{equation}
so dividing by the contribution from Fig.~\ref{fig:rho-acc-Tr} is appropriate at this order.

We first consider the limit in which the entropy of the environment $S_{\rm env}$ becomes much larger than that of the accessible degrees of freedom $S_{\rm acc}$. In this case, the slits in Fig.~\ref{fig:rho-acc-2} become small, and the sizes of the four connected components in Fig.~\ref{fig:rho-acc-2} all become comparable to that of the diagram in Fig.~\ref{fig:rho-acc}. Dividing the three diagrams by $Z^2$, we find that the contribution from the first diagram is of $O(1)$, while the contributions from the other, replica wormhole diagrams are suppressed by a factor of $1/Z$. This suppression is a result of the lack of wormhole connectivity when tracing out the environmental degrees of freedom.

The quantity $Z$ in Fig.~\ref{fig:rho-acc-Tr} is given by
\begin{align}
  Z &= \int\!{\cal D}q({\bf x})\, \bra{q({\bf x})} \left( \int\!{\cal D}A\, \ket{\Psi_A} \bra{\Psi_A} \right) \ket{q({\bf x})}
\nonumber\\
  &= \int\!{\cal D}q({\bf x})\, \bra{q({\bf x})} \eta \ket{q({\bf x})}.
\end{align}
While the integration runs over all configurations $q({\bf x})$, the rigging map ensures that only configurations inequivalent under $D$-dimensional diffeomorphisms contribute, up to the gauge volume, which cancels once a gauge is fixed. We thus find
\begin{equation}
  Z = \sum_I \int_{-\infty}^{\infty}\! dt\, e^{-i E_I t},
\end{equation}
where $t$ is a Lorentzian proper time, and $I$ runs over orthonormal basis states in $\hat{\cal H}_0$, the space of states physically inequivalent under diffeomorphisms (after appropriate regularization and for a given spacetime geometry). This yields
\begin{equation}
  Z = 2\pi \sum_I \delta(E_I) = 2\pi \delta(0) \sum_I 1,
\end{equation}
where we have used the fact that all the states have zero energy due to the Hamiltonian constraint. The $2\pi \delta(0)$ factor arises from the $t$ integral and is removed upon gauge fixing of global time translation, so we finally obtain
\begin{equation}
  Z \sim {\rm dim}\, \hat{\cal H}_0.
\label{eq:Z}
\end{equation}
We denote the logarithm of this quantity by $S_{\rm univ}$, which is approximately equal to $\ln {\rm dim}\,\hat{\cal H}_{\rm env}$ in this limit. Here, ${\rm dim}\,\hat{\cal H}_{\rm env}$ denotes the number of physically inequivalent environmental states (including boundary edge mode states), whose logarithm we refer to as $S_{\rm env}$.

We next consider the opposite limit in which $S_{\rm acc} \gg S_{\rm env}$. In this case, the slits nearly disconnect the geometries, so that all three diagrams in Fig.~\ref{fig:rho-acc-2} become similar, differing only by narrow necks connecting various rims of the cylinders. Therefore, as $S_{\rm env}$ decreases, we expect the wormhole contributions to become significant, eventually becoming $O(1)$, as in the fully disconnected case depicted in Fig.~\ref{fig:PXi2}. We thus expect that
\begin{equation}
  \overline{(\rho^{\rm acc}_{ij})^2} = \overline{\rho^{\rm acc}_{ij}}^2 \left[ 1 + O(e^{-S_{\rm env}}) \right].
\label{eq:precision}
\end{equation}
Note that this estimate accounts only for the leading topology and does not include contributions from more complex topologies (including but not limited to higher-genus surfaces).

The result in Eq.~(\ref{eq:precision}) can also be derived via a crude but more direct argument, as we now explain. The most significant difference between the first and the other two diagrams in Fig.~\ref{fig:rho-acc-2} is that, while the first term contains the square of the ``environmental partition function''
\begin{equation}
  Z_{\rm env} = \sum_a \int_{-\infty}^{\infty}\! dt\, e^{-i E_a t - \Gamma^2 t^2},
\label{eq:Z_env}
\end{equation}
the second and third terms contain only a single factor of it. (For the accessible degrees of freedom, there is no analogous enhancement from a sum over states.) Here, $a$ runs over orthonormal basis states in $\hat{\cal H}_{\rm env}$, and $\Gamma$ ($> 0$) reflects the fact that the environment is not a closed system, with $1/\Gamma$ representing the characteristic time scale over which environmental degrees of freedom remain in the system. The Gaussian damping term $-\Gamma^2 t^2$ in the exponent serves as a regularization of the $t$ integral.

The wormhole contributions from the second and third diagrams are therefore suppressed relative to that of the first diagram by the inverse of
\begin{equation}
  Z_{\rm env} = \frac{\sqrt{\pi}}{\Gamma} \sum_a e^{-\frac{E_a^2}{4\Gamma^2}}.
\end{equation}
While the $E_a$ may not be strictly zero, they are all close to zero due to the Hamiltonian constraints acting at each spatial point. The prefactor $\sqrt{\pi}/\Gamma$ arises from the $t$ integral and is removed upon fixing a gauge for global time translation. We thus find
\begin{equation}
  Z_{\rm env} \sim {\rm dim}\, \hat{\cal H}_{\rm env} = e^{S_{\rm env}},
\end{equation}
reproducing Eq.~(\ref{eq:precision}). This result will also be confirmed by a simple microscopic model discussed in the next section.

The rank of the reduced density operator $\overline{\rho^{\rm acc}}$ obtained by tracing out the environmental degrees of freedom is not $1$. The argument analogous to that used for (the appropriate extension of) Fig.~\ref{fig:PXi2} to show ${\rm rank}\, \rho = 1$ fails in the case of Fig.~\ref{fig:rho-acc-2}, where tracing over the environment removes the connectivity necessary to enforce a rank-$1$ structure. Together with the fact that the uncertainties arising from $\alpha$-microstates are suppressed by $e^{-S_{\rm env}}$, the elements $\overline{\rho^{\rm acc}_{ij}}$ of the reduced density matrix obtained from the path integral can be used to make robust predictions. The fractional error in these predictions is of order $e^{-\frac{1}{2}S_{\rm env}}$.

In fact, since the quantum gravitational state considered here incorporates the effects of standard time evolution, $\overline{\rho^{\rm acc}}$ is {\it einselected}~\cite{Zurek:1981xq,Zurek:2003zz}. Namely, for sufficiently large $S_{\rm env}$, $\overline{\rho^{\rm acc}}$ is approximately diagonal in the basis of pointer states $\ket{\psi}_{\rm acc}$ satisfying
\begin{equation}
  \mathop{\Tr}_{\rm env} \left[ U \left( \ket{\psi}_{\rm acc}\, {}_{\rm acc}\bra{\psi} \otimes \frac{\mathbb I}{{\rm dim}\, {\cal H}_{\rm env}} \right) U^\dagger \right] \approx \ket{\psi}_{\rm acc}\, {}_{\rm acc}\bra{\psi},
\end{equation}
where $U$ is the time evolution operator when the quantum gravitational state is interpreted via standard time evolution~\cite{DeWitt:1967yk,Page:1983uc}. Accordingly, physically meaningful questions are those phrased in this basis, with $\overline{\rho^{\rm acc}}$ interpreted as encoding classical relative probabilities. (For related discussions, see, e.g.,~\cite{Harlow:2025pvj,Nomura:2011rb}.)

The relative probability of obtaining the result $X$ in a physical measurement is given by
\begin{equation}
  P(X) = \sum_i \overline{\rho^{\rm acc}_{X_i X_i}},
\label{eq:pred}
\end{equation}
where the $X_i$ represent microstates in which the measurement has yielded result $X$, in the basis that $\overline{\rho^{\rm acc}}$ is approximately diagonal. Depending on the theory ${\cal I}$ and the question one asks, this may select a spacetime that does not give the largest contribution to the gravitational path integral.

We emphasize that we did {\it not} introduce any external ingredient beyond the elements of the original theory in extracting predictions. In particular, this does not require applying projection operators to the full universe state. Since the intrinsic uncertainty associated with this extraction is of order $e^{-\frac{1}{2}S_{\rm env}}$, the predictions become unreliable when $S_{\rm env}$ is small. In particular, it is not meaningful to pose physical questions involving all degrees of freedom in the entire universe.

\section{Microscopic Understanding in the Operator Formalism}
\label{sec:operator}

The mechanism described in this paper can be illustrated in the operator formalism using a simple model. We first discuss how uncertainties coming from $\alpha$-microstates are suppressed by tracing out inaccessible environmental degrees of freedom. We then show how this suppression enables making predictions in quantum gravity at a nonperturbative level, i.e., incorporating the effects of $\alpha$-microstates.

\subsection{Suppression of uncertainties from {\boldmath $\alpha$}-microstates}

We assume that a microscopic theory of quantum gravity contains degrees of freedom represented by $\alpha$-microstates $\ket{\alpha_\kappa}$ ($\in {\cal H}_\alpha$), which cannot be directly accessed at the semiclassical level. Suppose the unique quantum gravitational state of the universe corresponding to a given $\alpha$-microstate is
\begin{equation}
  \ket{\Omega_\kappa} = \sum_{n=1}^{e^{S_{\rm acc}}} \sum_{a=1}^{e^{S_{\rm env}}} c^\kappa_{na}\, \ket{\psi_n} \ket{\phi_a} \ket{\alpha_\kappa},
\end{equation}
where $\ket{\psi_n}$ and $\ket{\phi_a}$ represent orthonormal basis states of the constrained Hilbert spaces $\hat{\cal H}_{\rm acc}$ and $\hat{\cal H}_{\rm env}$, respectively, with $n$ and $a$ indexing these basis states.%
\footnote{
 Here we assume that the factorization of the Hilbert space ${\cal H}_0$ in Eq.~(\ref{eq:factor}) descends naturally to $\hat{\cal H}_0$. This amounts to ignoring correlations between states in $\hat{\cal H}_{\rm acc}$ and $\hat{\cal H}_{\rm env}$ due to the constraints.
}

The $\kappa$-dependent real coefficients $c^\kappa_{na}$ are treated as random variables and normalized such that $\inner{\Omega_\kappa}{\Omega_\kappa} = 1$, i.e., $\sum_n \sum_a (c^\kappa_{na})^2 = 1$. Note that the state $\ket{\alpha_\kappa}$ represents an $\alpha$-{\it microstate} and not an element of the baby universe Hilbert space ${\cal H}_{\rm baby}$, which emerges only when $\alpha$ is treated as a set of classical variables. Schematically, the entire set of microstates $\ket{\alpha_\kappa}$ corresponds to a single simultaneous eigenstate of baby-universe creation-annihilation operators of the form $a + a^\dagger$ in ${\cal H}_{\rm baby}$.

The density matrix of the universe, computed via the path integral, corresponds to an ensemble average over $\alpha$-microstates for which the bulk theory is regarded as having a well-defined coupling constant:
\begin{equation}
  \overline{\hat{\rho}_{(na)(mb)}} = \frac{1}{N_\alpha} \sum_{\kappa=1}^{N_\alpha}\, c^\kappa_{na} c^\kappa_{mb}.
\label{eq:av}
\end{equation}
Here, the sum runs over a complete set of orthonormal $\alpha$-microstates corresponding to a fixed value of the semiclassical parameter $\alpha$; the number of such states is denoted by $N_\alpha$. The hat on $\rho$ indicates that this is a density matrix in the {\it constrained} Hilbert space $\hat{\cal H}_0$. Similarly, the path-integral computation of the square of the matrix element, including replica wormholes, gives
\begin{equation}
  \overline{\hat{\rho}_{(na)(mb)}^2} = \frac{1}{N_\alpha} \sum_{\kappa=1}^{N_\alpha}\, (c^\kappa_{na})^2 (c^\kappa_{mb})^2.
\label{eq:sq-av}
\end{equation}

To proceed, we assume that the dependence of the coefficients $c^\kappa_{na}$ on $\kappa$ is random, such that for each fixed pair $(n,a)$, the $c^\kappa_{na}$ behave as independent random variables with zero mean:
\begin{equation}
  \overline{c^\kappa_{na}} = 0,
\qquad
  \overline{(c^\kappa_{na})^2} = \frac{1}{e^{S_{\rm univ}}},
\label{eq:stat}
\end{equation}
where the overlines denote averaging over $\kappa$, the second relation follows from the normalization of the state $\ket{\Omega_\kappa}$, and%
\footnote{
 This relation holds because, in this simple model, we have ignored correlations between $\hat{\cal H}_{\rm acc}$ and $\hat{\cal H}_{\rm env}$ arising from the constraints. In general, $S_{\rm univ}$ satisfies ${\rm max} \{ S_{\rm acc}, S_{\rm env} \} \leq S_{\rm univ} \leq S_{\rm acc} + S_{\rm env}$.
}
\begin{equation}
  S_{\rm univ} \equiv \ln {\rm dim}\, \hat{\cal H}_0 = S_{\rm acc} + S_{\rm env}.
\end{equation}
The square of the expectation value, $(\overline{\hat{\rho}_{(na)(mb)}})^2$, and the variance ${\rm Var}(\hat{\rho}_{(na)(mb)}) = \overline{\hat{\rho}_{(na)(mb)}^2} - (\overline{\hat{\rho}_{(na)(mb)}})^2$ are then given by
\begin{equation}
  (\overline{\hat{\rho}_{(na)(mb)}})^2 = \frac{1}{e^{2 S_{\rm univ}}},
\qquad
  {\rm Var}(\hat{\rho}_{(na)(mb)}) = \frac{2}{e^{2 S_{\rm univ}}}
\end{equation}
for $(na) = (mb)$, reproducing the relation in Eq.~(\ref{eq:PXi_Var}), and
\begin{equation}
  (\overline{\hat{\rho}_{(na)(mb)}})^2 = 0,
\qquad
  {\rm Var}(\hat{\rho}_{(na)(mb)}) = \frac{1}{e^{2 S_{\rm univ}}}
\end{equation}
for $(na) \neq (mb)$. These results assume Wick-like behavior for higher moments:\ a property satisfied by Gaussian and Haar-random ensembles. They imply that predictions based on the density matrix of the entire universe have $O(1)$ uncertainties, rendering them unreliable.

In contrast, the quantities corresponding to Eqs.~(\ref{eq:av}) and~(\ref{eq:sq-av}) for the reduced density matrix obtained after tracing the environmental degrees of freedom are
\begin{equation}
  \overline{\hat{\rho}^{\rm acc}_{nm}} = \frac{1}{N_\alpha} \sum_{\kappa=1}^{N_\alpha}\,  \sum_{a=1}^{e^{S_{\rm env}}} c^\kappa_{na} c^\kappa_{ma}
\end{equation}
and
\begin{equation}
  \overline{(\hat{\rho}^{\rm acc}_{nm})^2} = \frac{1}{N_\alpha} \sum_{\kappa=1}^{N_\alpha}\, \sum_{a,b=1}^{e^{S_{\rm env}}} c^\kappa_{na} c^\kappa_{ma} c^\kappa_{nb} c^\kappa_{mb},
\end{equation}
respectively. These lead to
\begin{equation}
  \overline{\hat{\rho}^{\rm acc}_{nm}} = 
  \begin{cases}
    \frac{1}{e^{S_{\rm univ}}} e^{S_{\rm env}} & \text{(for } n = m) \\
    0 & \text{(for } n \neq m),
  \end{cases}
\qquad\quad
  \overline{(\hat{\rho}^{\rm acc}_{nm})^2} = 
  \begin{cases}
    \frac{1}{e^{2 S_{\rm univ}}} (e^{2 S_{\rm env}} + 2 e^{S_{\rm env}}) & \text{(for } n = m) \\
    \frac{1}{e^{2 S_{\rm univ}}} e^{S_{\rm env}} & \text{(for } n \neq m),
  \end{cases}
\end{equation}
yielding
\begin{equation}
  {\rm Var}(\hat{\rho}^{\rm acc}_{n=m}) = 2 e^{-S_{\rm env}} (\overline{\hat{\rho}^{\rm acc}_{n=m}})^2,
\end{equation}
\begin{equation}
  \overline{(\hat{\rho}^{\rm acc}_{n \neq m})^2} = e^{-S_{\rm env}} (\overline{\hat{\rho}^{\rm acc}_{n=m}})^2.
\end{equation}
We thus find that for $S_{\rm env} \gg 1$, the uncertainty from the $\alpha$-microstate averaging is suppressed by a factor of order
\begin{equation}
  \frac{{\rm Var}(\hat{\rho}^{\rm acc}_{n=m})}{(\overline{\hat{\rho}^{\rm acc}_{n=m}})^2} \sim e^{-S_{\rm env}}.
\label{eq:uncertainty}
\end{equation}

The ensemble-averaged reduced density matrix becomes approximately proportional to the identity, up to corrections suppressed by $e^{-\frac{1}{2}S_{\rm env}}$:
\begin{equation}
  \overline{\hat{\rho}^{\rm acc}_{nm}} = \frac{1}{e^{S_{\rm acc}}} \left( \delta_{nm} + R_{nm} \right),
\label{eq:max-mixed}
\end{equation}
where $R_{nm}$ is a matrix whose elements, both diagonal and off-diagonal, are of order $e^{-\frac{1}{2}S_{\rm env}}$:
\begin{equation}
  R_{nm} \sim O(e^{-\frac{1}{2}S_{\rm env}}).
\end{equation}

This is consistent with the result of the path integral. While Eq.~(\ref{eq:max-mixed}) can be argued directly in the path integral, a simple way to see it is to consider the trace of the {\it unnormalized} average density matrix $\overline{\rho}$, which can be computed by the path integral in Fig.~\ref{fig:rho-acc-Tr}. To obtain the trace of its $n$-th power, $\overline{\rho}^n$---as opposed to $\overline{\rho^n}$---we need to cyclically connect $n$ copies of the cylinder representing $\overline{\rho}$ before closing the loop to form a torus. However, since the ``length'' of the torus is integrated over, the resulting path integral is the same as that in Fig.~\ref{fig:rho-acc-Tr}, giving
\begin{equation}
  \Tr \overline{\rho} = \Tr \overline{\rho}^n
\end{equation}
for all $n \in \mathbb{N}$. Considering $\Tr \overline{\rho} = Z \sim {\rm dim}\, \hat{\cal H}_0 = e^{S_{\rm univ}}$, as obtained in Eq.~(\ref{eq:Z}), the von~Neumann entropy of $\overline{\rho}$ is given by
\begin{equation}
  \lim_{n \rightarrow 1} \frac{1}{1-n} \ln \frac{\Tr \overline{\rho}^n}{(\Tr \overline{\rho})^n} = \ln (\Tr \overline{\rho}) = S_{\rm univ}.
\end{equation}
This suggests that the corresponding density matrix in the {\it constrained} Hilbert space $\hat{\cal H}_0$ is maximally mixed, $\overline{\hat{\rho}} = e^{-S_{\rm univ}} {\mathbb I}$, consistent with Eq.~(\ref{eq:max-mixed}).

\subsection{Making predictions in quantum gravity}

We have seen that taking the ensemble average over the $\alpha$-microsector renders the quantum gravity state maximally mixed in the constrained Hilbert space. Tracing out the environment makes it stable under fluctuations across the $\alpha$-microsector.

The implication of these observations is simple:\ when making physical predictions, we must treat all solutions to the constraint equations on an equal footing. By focusing on a subsystem of the universe---the accessible degrees of freedom---one can then make robust and reliable predictions.

Physical predictions are framed in a Hilbert space in which the Hamiltonian constraint is {\it not} imposed. After all, making predictions amounts to identifying physically allowed configurations (i.e., solutions to the Wheeler--DeWitt equation) within a ``fictitious'' space that contains all possible imaginary configurations, i.e., a Hilbert space in which the Hamiltonian constraint is not imposed (${\cal H}'_0$ or ${\cal H}_0$). This can be done by embedding the state $\overline{\hat{\rho}}$ or $\overline{\hat{\rho}^{\rm acc}}$ in the unconstrained Hilbert space. It is in this last step that the form of the action, ${\cal I}$ in Eq.~(\ref{eq:action}), becomes important, as realized by the path integral in Fig.~\ref{fig:rho-acc}.

Suppose the $n$-th element of the orthonormal basis in $\hat{\cal H}_{\rm acc}$ is embedded in ${\cal H}'_{\rm acc}$ as
\begin{equation}
  \ket{\psi_n} \bra{\psi_n} \;\rightarrow\; \sum_{i',j'} \overline{\rho'^{\rm acc}_{n, i'j'}} \ket{\psi_{i'}}' \bra{\psi_{j'}}',
\end{equation}
where the $\ket{\psi_{i'}}'$ are orthonormal basis states in ${\cal H}'_{\rm acc}$ which are naturally selected by the structure of locality encoded in the action ${\cal I}$~\cite{Nomura:2011rb}.%
\footnote{
 Here we treat the index $i'$ as discrete, but it can also be taken to be continuous, labeling the accessible portion of the configurations denoted by $A$ in Eq.~(\ref{eq:const-imp}). Similarly, $i$ in Eq.~(\ref{eq:Psi_n}) may label the accessible portion of the field configurations $q({\bf x})$ in Eq.~(\ref{eq:Psi_A}).
}
Note that since the map from ${\cal H}'_{\rm acc}$ to $\hat{\cal H}_{\rm acc}$ is not injective, an element of $\hat{\cal H}_{\rm acc}$ in general corresponds to a (possibly non-unique) mixed state in ${\cal H}'_{\rm acc}$. The corresponding state in ${\cal H}_{\rm acc}$ is then obtained by embedding this state into ${\cal H}_{\rm acc}$ through group averaging over the diffeomorphism group $G'$ of the $d$-dimensional spatial manifold:
\begin{equation}
  \int_{G'}\! dg'\, U'(g') \left( \sum_{i',j'} \overline{\rho'^{\rm acc}_{n, i'j'}} \ket{\psi_{i'}} \bra{\psi_{j'}} \right) U^{\prime -1}(g') \equiv \sum_{i,j} \overline{\rho^{\rm acc}_{n, ij}} \ket{\psi_i} \bra{\psi_j},
\label{eq:Psi_n}
\end{equation}
where $\ket{\psi_{i'}}$ denotes a representative state in ${\cal H}_{\rm acc}$ corresponding to $\ket{\psi_{i'}}'$ in ${\cal H}'_{\rm acc}$, and $U'(g')$ is a unitary representation of the $d$-dimensional diffeomorphism group $G'$. The $\ket{\psi_i}$ are orthonormal basis states in ${\cal H}_{\rm acc}$, also naturally selected by the locality of ${\cal I}$.

The (generally unnormalized) reduced density matrices relevant for making predictions in ${\cal H}'_{\rm acc}$ and ${\cal H}_{\rm acc}$ are then
\begin{equation}
  \overline{\rho'^{\rm acc}_{i'j'}} = \sum_{n=1}^{e^{S_{\rm acc}}} \overline{\rho'^{\rm acc}_{n, i'j'}},
\qquad
  \overline{\rho^{\rm acc}_{ij}} = \sum_{n=1}^{e^{S_{\rm acc}}} \overline{\rho^{\rm acc}_{n, ij}}.
\label{eq:rho-acc_H'-H}
\end{equation}
Note that while the reduced density matrix in the constrained Hilbert space---$\overline{\hat{\rho}^{\rm acc}_{nm}}$ in Eq.~(\ref{eq:max-mixed})---is approximately maximally mixed, the corresponding reduced density matrices in Eq.~(\ref{eq:rho-acc_H'-H}) are not. In fact, they are generally far from diagonal.

Diagonalizing $\overline{\rho'^{\rm acc}_{i'j'}}$ or $\overline{\rho^{\rm acc}_{ij}}$ yields the basis in which physical questions should be formulated. Since solutions to the Wheeler--DeWitt equation encode the outcome of time evolution in the conventional picture, we expect that the basis selected in this way coincides with the standard einselected basis. This ultimately stems from a key feature of the action ${\cal I}$: its locality in spacetime.

Finally, in the present analysis, we have traced out the environment before embedding the state into the unconstrained Hilbert spaces. However, the trace could instead be performed after the embedding, as was the case in the path integral. This does not affect the result, under the current simplifying assumption that correlations between $\hat{\cal H}_{\rm acc}$ and $\hat{\cal H}_{\rm env}$ are neglected.

\section{Conclusions and Discussion}
\label{sec:concl}

We have examined how robust physical predictions can emerge in nonperturbative quantum gravity in a closed Lorentzian universe, where the Hilbert space is believed to be one-dimensional and real for each $\alpha$-sector. This structure appears to obstruct the conventional probabilistic interpretation of quantum mechanics through projections onto a basis of states.

Rather than resolving this tension by introducing additional external ingredients or observers, we argued that quantum gravity itself, properly understood, contains all the necessary elements to recover classical observables and probabilistic outcomes. Our results suggest that the apparent obstructions to probabilistic predictions in quantum gravity---stemming from the structure of the Hilbert space and the absence of external observers---are naturally resolved by the physical fact of partial observability. When the inaccessible degrees of freedom are traced out, the resulting reduced density matrix can support classical predictivity, with uncertainties that are exponentially suppressed by the entropy of the environment. In this view, physical predictions are inherently relational and confined to accessible subsystems, highlighting the limited scope of questions that can meaningfully be posed in a closed universe.

To move toward more realistic cosmological settings, we conclude by discussing two aspects of the framework presented in this paper. First, since physical questions are framed locally in the accessible degrees of freedom, the predictions do not seem to depend on the global spacetime; in particular, as anticipated in Refs.~\cite{Nomura:2011dt,Nomura:2011rb,Bousso:2008hz,Garriga:2012bc,Friedrich:2022tqk}, probabilities do not seem to be weighted by the global volume. Second, the quantum gravitational state considered in this paper is maximally mixed over the set of states satisfying the physical constraints. However, it is possible that to obtain a realistic cosmology, we may need to impose some further condition on the wavefunction, for example some sort of normalizability condition~\cite{Nomura:2012zb}, in which case the projection onto the quantum gravitational state will be modified as
\begin{equation}
  \int\!{\cal D}A\, \ket{\Psi_A} \bra{\Psi_A} \;\;\rightarrow\; \int\!{\cal D}A\, {\cal D}A'\, {\cal F}(A,A')\, \ket{\Psi_A} \bra{\Psi_{A'}},
\end{equation}
where ${\cal F}(A,A') = {\cal F}^*(A',A)$ is a functional of $A$ and $A'$, determined by the additional condition. In the conventional multiverse picture based on the string landscape, it is expected that the spatial curvature of our own universe is negative~\cite{Kleban:2012ph,Guth:2012ww}. It would be interesting to see whether a multiverse scenario with positive spatial curvature can be constructed using the framework presented in this paper.

The findings of this paper offer a new perspective on the emergence of classicality in quantum gravity, and we expect that further exploration of this viewpoint will shed additional light on the nature of observables and dynamics in the quantum universe.

\vspace{0.3cm}
\flushleft{\bf\large Note Added:}

While this manuscript was being completed, Refs.~\cite{Blommaert:2025bgd,Chen:2025fwp} appeared, which discuss related issues.

\section*{Acknowledgments}

This work was supported in part by MEXT KAKENHI Grant Number JP25K00997. The work of Y.N. was also supported by the U.S. Department of Energy, Office of Science, Office of High Energy Physics under QuantISED award DE-SC0019380 and contract DE-AC02-05CH11231.

\end{document}